# PoolRank: Max/Min Pooling-based Ranking Loss for Listwise Learning & Ranking Balance

Zhizhong Chen
Brown University
Providence, RI, USA

Carsten Eickhoff
Brown University
Providence, RI, USA

## ABSTRACT

Numerous neural retrieval models have been proposed in recent years. These models learn to compute a ranking score between the given query and document. The majority of existing models are trained in pairwise fashion using human-judged labels directly without further calibration. The traditional pairwise schemes can be time-consuming and require pre-defined positive-negative document pairs for training, potentially leading to learning bias due to document distribution mismatch between training and test conditions. Some popular existing listwise schemes rely on the strong pre-defined probabilistic assumptions and stark difference between relevant and non-relevant documents for the given query, which may limit the model potential due to the low-quality or ambiguous relevance labels. To address these concerns, we turn to a physics-inspired ranking balance scheme and propose PoolRank, a pooling-based listwise learning framework. The proposed scheme has four major advantages: (1) PoolRank extracts training information from the best candidates at the local level based on model performance and relative ranking among abundant document candidates. (2) By combining four pooling-based loss components in a multi-task learning fashion, PoolRank calibrates the ranking balance for the partially relevant and the highly non-relevant documents automatically without costly human inspection. (3) PoolRank can be easily generalized to any neural retrieval model without requiring additional learnable parameters or model structure modifications. (4) Compared to pairwise learning and existing listwise learning schemes, PoolRank yields better ranking performance for all studied retrieval models while retaining efficient convergence rates.

## CCS CONCEPTS

• **Information systems** → **Retrieval models and ranking**; **Evaluation of retrieval results**; *Search engine architectures and scalability.*

## KEYWORDS

Listwise Ranking Loss, Pooling Technique, Ranking Balance, Partial Relevance, Learning to Rank, Information Retrieval

## 1 INTRODUCTION

Deep learning has been successfully applied to many problems, from language understanding to multilingual translation, from information extraction to text matching and retrieval. Various neural ranking models [4, 8, 11, 18, 19, 27, 29, 30] have been proposed to describe the relationship between a query and the candidate documents. These models learn rich representations that summarize all salient matching information. The representations, in turn, are used to compute a ranking score that quantifies the relevance level between the given query and document.

Unlike loss functions in classification and regression settings, the optimization goal of neural retrieval models is *ad hoc* due to the non-differentiable nature of the ranking problem. Most neural retrieval models are trained in pairwise fashion. RankNet [2] and LambdaRank [1] have solved many real-world ranking problems successfully. Recently, pairwise margin ranking loss [12, 26] has been a popular choice for many neural retrieval models [4, 8, 11, 16, 18, 19, 30]. However, in most realistic applications, the number of non-relevant documents is likely to be much larger than that of truly relevant documents. The artificially selected document pairs for pairwise learning may misrepresent the true document relevance distribution in the underlying collection. Hence, various listwise learning schemes such as ListNet [3], ListMLE [28], ApproxNDCG [22], and SoftRank [25] have been proposed to explore the relative ordering among different document candidates directly based on strong pre-defined probabilistic assumptions or sophisticated mathematical approximation.

The reliability of relevance labels and the stark content contrast between relevant documents and non-relevant candidates are the keys to validate the reasoning of those existing pairwise/listwise learning schemes and to ensure their significant performance. Nonetheless, many human-judged labels can be insufficient and erroneous to identify all the possible truly relevant documents from a huge candidate collection in many realistic applications. The high labor cost and the availability of skillfully trained human relevance judge further prevent academic researchers from the independent label calibration on a large scale. Moreover, the nuance among certain documents and the convoluted query/document content may lead to partial relevance or even arguable disagreement among different human judges.

In this paper, we introduce the profound concept in physics, mechanical equilibrium, to information retrieval and propose a new pooling-based listwise learning framework, under which partial relevance and potentially low-quality labels are considered directly via min/max-pooling techniques to aid model training. A min-pooling layer aims to select the best non-relevant document candidates at the local level. A max-pooling layer is used to exploit the partial relevance information from potential partially relevant, yet non-relevant-labeled candidates carefully based on the prediction comparison. The four components of PoolRank then preserve an effective ranking balance for the selected minimum/maximum candidates to regulate model predictions so that the scores of the highly non-relevant documents remain at the lower range, while keeping moderate score level for the partially relevant documents. We refer to this scheme as PoolRank.

We apply PoolRank to three neural retrieval models: ConvKNRM [30], KNRM [4] and DRMM [11]. To demonstrate the effectiveness, robustness and scalability of our framework, we compare PoolRank with the traditional pairwise/listwise learning approaches on the MS MARCO collection [20]. Finally, to further validate the impact of partial relevance, a limited set of news collection data is used to train these models with the restricted access to the relevance labels.

The major novel contributions of this paper are (1) Formulation of a new pooling-based listwise learning framework using pooling techniques to incorporate the concept of "Ranking Balance", inspired by mechanical equilibrium in physics. (2) The direct consideration of partial relevance within PoolRank to tackle the label quality issue automatically. (3) Empirical verification of generalizability, robustness, scalability and effectiveness of the proposed listwise framework.

The remainder of this paper is organized as follows. Section 2 discusses related work on pairwise and listwise learning approaches. Section 3 formally introduces the mathematical formulation of PoolRank framework. Section 4 describes our experimental design, implementation settings and a performance comparison between PoolRank and existing pairwise/listwise learning approaches. An ablation study of the individual PoolRank loss components is presented in Section 5. In Section 6, we study the effect of pooling size – a critical parameter in our framework. Finally, Section 7 concludes with a discussion of key findings and future directions of inquiry.

## 2 BACKGROUND AND RELATED WORK

*Learning to rank* (L2R) is a supervised retrieval setting for which many neural network architectures have been proposed. Neural retrieval models exploit local matching signals between query and document based on dense numerical vector representations, and then use neural network architectures to estimate the relevance level for document ranking. Examples include DeepMatch [16], KNRM [4], ConvKNRM [30], MatchPyramid [18], DRMM [11], PACRR [8], and DeepRank [19]. According to the formulation in the original papers, these models are trained in a pairwise manner, where positive-negative document pairs are used to train a pairwise margin ranking loss $L$

$$L(q, d^+, d^-) = max(0, 1 - s(q, d^+) + s(q, d^-)) \qquad (1)$$

where $s(q, d)$ is the ranking score between query $q$ and document $d$, and document $d^+$ is more relevant than $d^-$ for query $q$.

Similarly, RankNet [2] computes the probability that the ranking score of a relevant document is higher than that of a non-relevant document and then optimizes a cross entropy loss. However, document pair generation can be computationally costly, as the number of theoretically possible document pairs is huge. In most realistic applications, the number of non-relevant documents is likely much larger than that of truly relevant documents. Randomly selected document pairs during the pairwise training period may misrepresent the true document relevance distribution in the underlying collection. Selective negative sampling methods have been proposed to train models [6, 10, 17]. Recently, Cohen *et al.* [5] have applied reinforcement learning to optimize a policy over a set of sampling functions to select non-relevant documents for model training.

To address the potential drawbacks of pairwise learning, listwise learning schemes have been proposed. Listwise learning jointly evaluates multiple documents per query during the training period. Due to the non-differentiable nature of the ranking problem, such loss formulations can become non-trivial. A number of methods have been proposed to find the optimal ordering of the whole list of document candidates to aid training. There are two major trends in listwise learning: 1) Minimizing an axiomatic loss function which is derived based on the nature of ranking problems, or, 2) Optimizing concrete ranking metrics, such as nDCG [9], MRR [23], or MAP [15], directly.

Probability-based listwise loss functions showed excellent performance in the past, especially ListNet [3], and ListMLE [28] have been highly successful. ListNet [3] employs a softmax function to transform both model generated ranking scores and human-generated judgments into probability distributions. Then, a cross entropy loss is applied to jointly train the models based on a list of documents. On the other hand, ListMLE [28] treats the ranking problem as a sequential selection process, maximizing the probability of the observed permutation directly by minimizing the negative sum of log likelihoods with respect to all training queries.

Metrics-based optimization gives us an intuitive alternative to listwise learning. Since ranking metrics are usually non-differentiable, various sophisticated mathematical approximation methods have been formulated to generate smooth surrogate functions, such as ApproxNDCG [22] and SoftRank [25]. Moreover, the recently proposed IRGAN [27] combines adversarial learning and reinforcement learning to solve information retrieval problems by jointly utilizing two competing neural networks, instead of constructing a novel loss function.

Probability-based listwise learning approaches presume certain distributions to be valid and suitable for explaining the relative ordering within the global list of candidate documents, which in fact is difficult to justify universally for different problem contexts. Conversely, the approximated target of metrics-based optimization is often not convex. As a result, the training process can get stuck in local minima during gradient-based optimization steps.

As an alternative, we propose PoolRank, a pooling-based listwise learning framework. PoolRank avoids assuming that scores or relevance labels follow any particular distribution. It employs the classic pooling technique to achieve ranking balance for non-relevant-labeled documents. Via this effective model prediction-based selection technique, PoolRank can squeeze more salient information into model and accelerate the learning process.

## 3 POOLING-BASED LISTWISE RANKING LOSS

The innate ability of listwise learning to compare a list of candidates simultaneously to find the relative ordering for the final prediction makes it more consistent with the comparison-based ranking nature of information retrieval. Hence, it usually provides much more significant and robust performance than pairwise learning. However, the existing listwise approaches treat the given candidate list as a whole without further inspections.

This section describes PoolRank, a novel listwise learning framework relying on the pooling technique and the constructed ranking balance. The proposed ranking loss consists of 4 parts: min-pooling margin ranking loss, min-max distance loss, max-pooling regularization and target regularization. Inspired by the concept of mechanical equilibrium in physics, different candidates in the given list are treated as "particles" in a special "physical system", where they interact with each other during model training. Once the system stabilizes, their predicted ranking scores can reveal their true relevance levels with respect to the given query. Four loss components within PoolRank can be regarded as four various potential energies to preserve an effective ranking balance for the score levels of the selected document candidates. During back propagation optimization, the derivative of each loss component is utilised as a mechanical force to regulate the corresponding ranking scores. The overall structure of PoolRank is displayed in Figure 1.

### 3.1 Candidate Assumption

Given a query $q$, our listwise training data contains $N$ known relevant documents $d_1^+, ..., d_N^+$ and $M$ documents $d_1^-, d_2^-, ..., d_M^-$ which are marked as non-relevant according to our annotators. The foundation of PoolRank relies on the intuitive assumption that there are three types of candidates: 1) Relevant documents, 2) Truly non-relevant documents, 3) Partially Relevant, yet non-relevant-labeled documents. In many realistic applications, human-judged labels can be insufficient and erroneous to identify all the possible truly relevant documents consistently from a huge candidate collection. As a result, even though many documents are labeled as negative, there is still a gradation of relevance which induces a relative ordering within the document list. The range of partial relevance often is query-dependent, model-dependent and relatively distinguishable within document candidates. How to effectively select those partially relevant candidates is the key to further boost the learning performance.

### 3.2 Pooling-based Candidate Selection

PoolRank employs the classic pooling technique to group non-relevant-labeled documents and to identify partially relevant candidates within each pooling group based on model predictions automatically. Given any neural retrieval model that outputs a ranking score $s(q, d)$ between a query $q$ and a document $d$, a pair of min/max-pooling layers, with the same pooling window size $\kappa$, is used to select the most distinct non-relevant-labeled documents from $s(q, d_1^-), s(q, d_2^-), ..., s(q, d_M^-)$. Within each pooling window, the min-pooling layer selects the most obviously non-relevant documents with minimum scores: $s(q, d_{\delta_1}^-), s(q, d_{\delta_2}^-), ..., s(q, d_{\delta_m}^-)$, and the max-pooling layer chooses the most likely partially relevant documents with maximum scores: $s(q, d_{\Delta_1}^-), s(q, d_{\Delta_2}^-), ..., s(q, d_{\Delta_m}^-)$, where $m = \lceil \frac{M}{\kappa} \rceil$.

Within each pooling window with scores of $s_1^-, s_2^-, ..., s_\kappa^-$, the max-pooling layer $P_{max}$ is indicated as

$$P_{max}(s_1^-, s_2^-, ..., s_\kappa^-) = max(s_1^-, s_2^-, ..., s_\kappa^-) = s_\Delta^- \quad (2)$$

The partial derivative with respect to each score is

$$\frac{\partial P_{max}}{\partial s_\Delta} = 1 \quad \frac{\partial P_{max}}{\partial s_i} = 0, \ i \neq \Delta \quad (3)$$

This means the gradient-based optimization update from the next layer will only be passed back to the query-document pairs with the local maximum scores within each pooling window during back propagation. The same logic applies to the min-pooling layer. Hence, the max/min-pooling technique can be regarded as a document-level drop-out operation to ignore certain mediocre candidates to aid model training.

### 3.3 Min-Pooling Margin Ranking Loss

Compared to other candidates within each pooling window, the selected minimum-score documents are more likely to be truly non-relevant to the query. Unlike ListNet[3] and ListMLE[28], the traditional pairwise margin ranking loss avoids the dependence on the strong pre-defined probabilistic assumptions and yields simple yet effective forces to drive up the positive scores and to push down the non-relevant scores, which stabilizes the ranking balance on the minimum candidates.

Since the number of relevant documents $N$ is relatively much smaller than $M$ and for some collections and tasks may even be $N = 1$, we use the average of positive scores in the margin ranking loss, $\bar{s}(q, d^+) = \frac{1}{N} \sum_{i=1}^{N} s(q, d_i^+)$. We believe that when there are abundant positive documents available, compared to the size of $m$, multiple document lists can be constructed separately to contain a single relevant document within each list to lead similar performance, instead of using the average positive score. Thus, the min-pooling margin ranking loss can be written as:

$$L_{min} = \frac{1}{m} \sum_{i=1}^{m} max(0, 1 - \bar{s}(q, d^+) + s(q, d_{\delta_i}^-)) \quad (4)$$

### 3.4 Min-Max Distance Loss

During early training stages, parameters within neural retrieval models change dramatically from initialization and their convergence level is reasonably low, which leads to comparative inconsistency between the relative ranking order of $s(q, d)$ obtained from model predictions and the actual relevance ranking of those documents. As a result, $L_{min}$ alone at the early stage may force the scores of some partially relevant documents down to -1, due to the pooling selection quality.

Ideally, the scores of partially-relevant documents should fall relatively nearer the score range of relevant documents intrinsically. Hence, a reliable neural retrieval model should assign relatively high scores. To enforce this axiom, min-max distance loss $L_{min/max}$ is then proposed to prevent the selected minimum scores from becoming too small. In this way, $L_{min/max}$ and $L_{min}$ can achieve a reasonable score balance for the documents selected from the min-pooling layer. The min-max distance loss is formulated as

$$L_{min/max} = \frac{1}{m} \sum_{i=1}^{m} (s(q, d_{\Delta_i}^-) - s(q, d_{\delta_i}^-))^2 \quad (5)$$

### 3.5 Max-Pooling Regularization

Due to the convergence level of neural retrieval models and given candidate list quality, the selected maximum documents may contain both truly non-relevant documents and partially relevant candidates. Max-pooling regularization $L_{max}$ is then formulated to prevent those truly non-relevant documents within maximum candidates from becoming too big. Most neural retrieval models output ranking scores via a final tanh layer. The range of ranking scores is $[-1, 1]$. The max-pooling regularization term is then written as:

$$L_{max} = \frac{1}{m} \sum_{i=1}^{m} (s(q, d_{\Delta_i}^-) + 1)^2 \qquad (6)$$

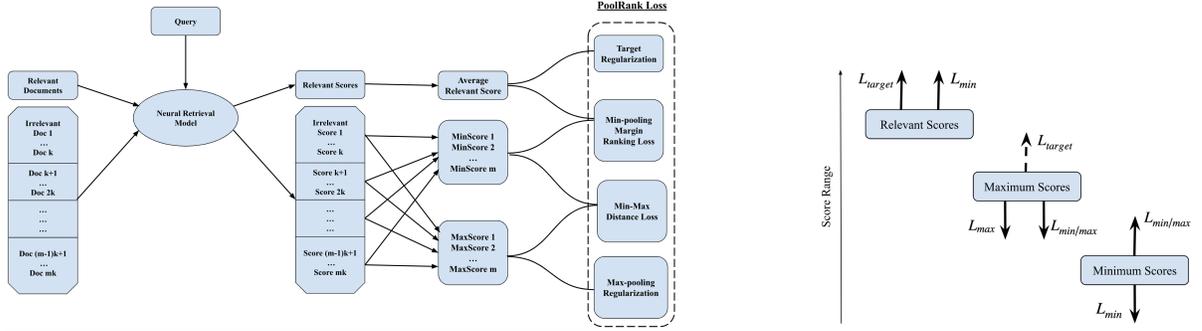

**Figure 1: Left: The overall structure of the Pooling-based Listwise Learning Framework, PoolRank. RelevantScores are the ranking scores of the positive relevant documents. MinScores and MaxScores are non-relevant document scores selected from the min/max-pooling layers, respectively. Right: An illustration of the Ranking Balance formed by the four components within PoolRank. The arrows indicate the effect direction of each loss component on the given score group. The dashed arrow indicates the implicit impact of $L_{target}$ on the maximum candidates.**

### 3.6 Target Regularization

Similarly, we further add a target regularization term $L_{target}$ to boost scores for relevant documents.

$$L_{target} = (1 - \bar{s}(q, d^+))^2 \qquad (7)$$

Due to the implicit resemblance between relevant documents and partially relevant candidates, $L_{target}$ also exerts an indirect impact on the partially relevant documents within maximum candidates. As a result, $L_{min/max}$, $L_{max}$ and $L_{target}$ together establish an subtle ranking balance for the maximum scores.

### 3.7 Listwise Ranking Loss & Ranking Balance

Our final listwise learning framework is formulated as a multi-task learning scheme via the weighted summation of the four components above:

$$L_{PoolRank} = c_1 L_{min} + c_2 L_{min/max} + c_3 L_{max} + c_4 L_{target} \qquad (8)$$

Wang *et al.* [27] introduce adversarial learning to information retrieval based on the competition between a generative and a discriminative model. Their proposed generative model uses policy gradient based reinforcement learning to select candidate documents. However, the Markovian nature and the predominantly online learning paradigm can prevent the algorithm from performing well in realistic scenarios. Especially with potential shallowly annotated relevance labels, reinforcement learning may further deteriorate under the circumstances of the misguided relevance feedback. When there are abundant document candidates, IRGAN can be time-consuming and sensitive to the initial setting, due to its sequential decision making. Additionally, IRGAN requires the tailored design of neural networks for both generative model and discriminative model, which does not allow for seamless integration into arbitrary existing neural retrieval models.

Similar to the core idea of competition within adversarial learning, ranking balance is the key concept to ensure the success of PoolRank and to tackle the partial relevance problem within annotated labels, which is illustrated in Figure 1. $L_{min}$ effectively drives up the relevant scores and pushes down the minimum scores. $L_{min/max}$ creates a contraction force to prevent minimum scores from being too small. As a result, $L_{min}$ and $L_{min/max}$ create a ranking balance for the minimum scores to remain at a reasonably lower level. On the other hand, since some of the maximum candidates may still be reasonably non-relevant, $L_{max}$ further pushes down the corresponding scores of truly non-relevant documents among the maximum candidates. The partial similarity between positive candidates and partially relevant candidates enforces $L_{target}$ to enhance the maximum scores indirectly. Hence, $L_{min/max}$, $L_{max}$ and $L_{target}$ together establish an subtle ranking balance for the maximum scores.

Formulating our goal as a multi-task learning problem, the resulting listwise learning framework offers a number of noteworthy desirable properties. First of all, since our listwise ranking loss is solely based on min-pooling and max-pooling operations, there is no need for introducing additional learnable parameters, making model training less complex and reducing the risk of overfitting. The document-level drop-out effect, induced by the pooling layers, further ignores certain atypical candidates to accelerate model training. The only requirement towards our listwise learning framework is for the range of output ranking scores to fall in the $[-1, 1]$ interval. Various neural retrieval models can be easily fit into our framework. Also, pooling techniques can extract relative ordering information at the local level without sorting. This property preserves the differentiability of the whole end-to-end ranking framework, which is essential for effective back propagation. Moreover, as an efficient selection process, it is easy and straightforward to apply our listwise learning approach to datasets with different sizes and scales.

## 4 EXPERIMENT DESIGN

As the theoretical aspects of our pooling-based listwise learning seem promising, we conduct multiple empirical experiments to demonstrate PoolRank's performance compared to existing pairwise and listwise learning schemes based on various neural retrieval models.

### 4.1 Retrieval Models

We consider the following neural ranking models in our performance comparison:

**KNRM** [4] is a kernel based neural model for document ranking. It computes word-level similarities via word embeddings of query and document terms and uses Gaussian kernel pooling techniques to extract multi-level soft-match features. Its final ranking score is then given via a learning-to-rank layer that combines those features.

**ConvKNRM** [30] is a convolutional kernel-based extension of KNRM model. It utilizes Convolutional Neural Networks (CNN) [14] to compose adjacent words' embeddings into n-gram embeddings. Similarly, kernel pooling and learning-to-rank layers are used for the final ranking score computation.

**DRMM** [11] is a deep relevance matching model for ad-hoc retrieval. It employs matching histogram mapping to summarise word-level similarities between query word embeddings and document word embeddings. A term gating network, which is based on word's Inverse Document Frequency (IDF), is used to help neural network to generate ranking score.

We keep the same ranking model structure so that performance between pairwise learning and listwise learning is comparable. We evaluate and compare our models in terms of the total training time and three common ranking metrics: the Mean Reciprocal Rank (MRR), the Normalized Discounted Cumulative Gain (nDCG) and the Mean Average Precision (MAP) [9, 15, 23].

### 4.2 Model Implementation Setting

We implement all neural retrieval models in tensorflow[1]. Our GloVe word embeddings are trained on the Common Crawl dataset [2]. We use an Adam optimizer [13] with 0.0001 learning rate to train all models. For both KNRM and ConvKNRM models, we set the number of Gaussian kernels to 11 with one harvesting exact matches, using mean value $\mu_0 = 1.0$ and standard deviation $\sigma_0 = 10^{-3}$, and ten capturing soft matches, using mean values spaced evenly in the range of $[-1, 1]$, $\mu_1 = 0.9, \mu_2 = 0.7, ..., \mu_{10} = -0.9$, and standard deviation $\sigma_1 = 0.1$. For ConvKNRM specifically, we set the number of CNN filters to 128 and consider only n-gram lengths of 1, 2, 3. For DRMM, we use LogCount-based histogram (LCH) mapping and set histogram mapping dimension to 30.

### 4.3 Performance on MS MARCO

The MS MARCO (Microsoft Machine Reading Comprehension) passage ranking dataset [20] is a large-scale dataset which contains nearly 550,000 unique queries from the Bing search engine and more than 8.8 million unique passages. For each query, BM25 [24] is used to produce the top 1000 corresponding passages and human judges annotated any potentially matching passage. Most queries have only one positive relevant passage. Since the difference among some corresponding passages can be difficult to judge, some passages may be erroneously marked as non-relevant even though they are relevant to the query to some limited extent. The collection is comprised of the top1000 training dataset and the top1000 development dataset. We follow common practice in using 20% of the development set (1,396 queries) for model validation, and the remaining 80% (5,583 queries) as test data for both pairwise learning and listwise learning. The maximum word length of queries and documents is set to 15 and 150, respectively.

[1] https://www.tensorflow.org/
[2] https://commoncrawl.org/

The major advantage of listwise learning comes from its ability to compare a list of candidates simultaneously. Therefore, we choose 4 different list sizes to examine the its impact on the final performance. All truly relevant candidates are always included. Then we randomly sample 50, 100, 150, 200 negative candidates from the top 1000 corresponding passages for each query during training, respectively. To make the training condition comparable between pairwise learning and listwise learning, we select one positive passage and pair it with all the negative passages in the corresponding candidate list to create positive-negative passage pairs for each query during pairwise learning. Validation dataset and test dataset are the same for all the candidate settings, where there are approximately 1000 documents associated with each query. The final test performance is based on the relative ranking performance among those 1000 documents for each query in the test dataset.

Since MS MARCO is a raw text dataset, we use GloVe embeddings [21] with 300 dimensions as our word embedding initialization. Normal random initialization is used when the words or tokens are not found in GloVe collection. We also replace low frequency words/tokens by <UNK> tokens if their appearance frequency is less than 10 among all the queries and passages [7]. Any gains incurred from replacing the representation layer by more powerful transformer-based alternatives can be expected to equally affect all model architectures and are left out of the scope of this work for simplicity.

The basic model settings above are chosen for both pairwise learning and listwise learning to ensure $s(q, d)$ to be consistent. For pairwise learning, we train our models with different batch sizes and choose the optimal ones for each model. For listwise learning, the batch size is set to 4. For each list size, we train our model with different pooling sizes and report the best-performing ones.

For simplicity, the mixture parameters of PoolRank are set to $c_1 = 0.5, c_2 = 1, c_3 = 0.5, c_4 = 1$ for all the models and list sizes, which is based on the validation performance of ConvKNRM model with list size of 50. Since ranking balance is the cornerstone of PoolRank, these weight parameters control the magnitudes of different forces to regulate maximum/minimum scores. Hence, we do believe that model-dependent and data-dependent fine tuning can further strengthen PoolRank performance. The general effect of each component weight is discussed in Sec 5.

To enable fair comparisons with PoolRank, for pairwise learning, we use pairwise margin ranking loss and RankNet as the key baselines. For listwise learning, the probability-based ListMLE and ListNet are selected, along with the metrics-based ApproxNDCG. The test set performance for different training document levels (Top-50, Top-100, Top-150, Top-200) is shown in Table 1.

Table 1: Test Performance on MS MARCO with different list sizes. $\kappa$ refers to the best selected pooling window size in PoolRank. ¶ indicates statistically significant improvements over all the other losses. § indicates full statistically significant improvement except ListNet. ‡ indicates full statistically significant improvement except ListMLE. † indicates full statistically significant improvement except ListNet and ListMLE.

| Model | Top 50 | | | | | Top 100 | | | | | Top 150 | | | | | Top 200 | | | | |
|---|---|---|---|---|---|---|---|---|---|---|---|---|---|---|---|---|---|---|---|---|
| | MRR | nDCG | MAP | Time | $\kappa$ | MRR | nDCG | MAP | Time | $\kappa$ | MRR | nDCG | MAP | Time | $\kappa$ | MRR | nDCG | MAP | Time | $\kappa$ |
| DRMM_Margin | 0.14723 | 0.27916 | 0.14414 | 7.96h | | 0.14561 | 0.27566 | 0.14187 | 6.64h | | 0.14357 | 0.27621 | 0.14078 | 13.49h | | 0.14880 | 0.28138 | 0.14588 | 19.89h | |
| DRMM_RankNet | 0.14804 | 0.28004 | 0.14490 | 6.09h | | 0.15180 | 0.28396 | 0.14850 | 15.11h | | 0.14745 | 0.27902 | 0.14457 | 18.92h | | 0.15294 | 0.28528 | 0.14980 | 21.68h | |
| DRMM_ListNet | 0.17121 | 0.30538 | 0.16801 | **4.71h** | | 0.19021 | 0.32389 | 0.18678 | 7.69h | | 0.19856 | 0.33256 | 0.19501 | 10.44h | | 0.19860 | 0.33139 | 0.19537 | 12.73h | |
| DRMM_ListMLE | 0.20288 | 0.33608 | 0.19962 | 13.8h | | 0.20101 | 0.33396 | 0.19811 | 14.24h | | 0.19984 | 0.33263 | 0.19691 | 26.78h | | 0.19980 | 0.33308 | 0.19688 | 32.71h | |
| DRMM_nDCG | 0.14735 | 0.27939 | 0.14472 | 10.50h | | 0.14609 | 0.27870 | 0.14334 | 15.03h | | 0.15094 | 0.28283 | 0.14815 | 18.05h | | 0.15020 | 0.28204 | 0.14770 | 21.70h | |
| DRMM_PoolRank | **0.20467‡** | **0.33806‡** | **0.20114‡** | 7.75h | 15 | **0.20231‡** | **0.33414‡** | **0.19863‡** | 6.49h | 10 | **0.20182†** | **0.33355†** | **0.19818†** | 10.16h | 10 | **0.20219†** | **0.33363†** | **0.19864†** | 11.05h | 10 |
| KNRM_Margin | 0.20885 | 0.34811 | 0.20475 | 6.41h | | 0.20520 | 0.34385 | 0.20130 | 5.16h | | 0.20403 | 0.34307 | 0.20057 | 5.10h | | 0.19115 | 0.32964 | 0.18754 | 21.23h | |
| KNRM_RankNet | 0.20025 | 0.33908 | 0.19669 | 8.97h | | 0.20713 | 0.34483 | 0.20403 | 12.18h | | 0.19764 | 0.33511 | 0.19396 | 6.34h | | 0.18369 | 0.32191 | 0.18046 | 6.87h | |
| KNRM_ListNet | 0.23719 | 0.37184 | 0.23378 | **2.64h** | | 0.24090 | 0.37560 | 0.23742 | 4.29h | | 0.24080 | 0.37562 | 0.23726 | **4.56h** | | 0.24348 | 0.37741 | 0.23983 | **3.96h** | |
| KNRM_ListMLE | 0.23427 | 0.36926 | 0.23079 | 7.96h | | 0.23010 | 0.36460 | 0.22647 | 9.51h | | 0.22951 | 0.36427 | 0.22635 | 16.96h | | 0.22264 | 0.35765 | 0.21939 | 12.76h | |
| KNRM_nDCG | 0.23077 | 0.36686 | 0.22721 | 5.54h | | 0.23418 | 0.36975 | 0.23088 | 6.28h | | 0.23443 | 0.36873 | 0.23119 | 5.76h | | 0.23387 | 0.36837 | 0.23068 | 6.56h | |
| KNRM_PoolRank | **0.24572¶** | **0.38012¶** | **0.24174¶** | 3.84h | 25 | **0.24722¶** | **0.38147¶** | **0.24305¶** | 4.22h | 30 | **0.24615¶** | **0.38059¶** | **0.24207¶** | 5.07h | 30 | **0.24681§** | **0.38098§** | **0.24276§** | 5.78h | 40 |
| Conv-KNRM_Margin | 0.23255 | 0.36923 | 0.22828 | 20.51h | | 0.23687 | 0.37295 | 0.23230 | 42.94h | | 0.24104 | 0.37828 | 0.23703 | 43.64h | | 0.23676 | 0.37394 | 0.23288 | 55.32h | |
| Conv-KNRM_RankNet | 0.18307 | 0.32358 | 0.17917 | 34.22h | | 0.20517 | 0.34217 | 0.20116 | 25.06h | | 0.22667 | 0.36459 | 0.22292 | 39.35h | | 0.22122 | 0.35864 | 0.21738 | 60.99h | |
| Conv-KNRM_ListNet | 0.27928 | 0.41123 | 0.27508 | **6.67h** | | 0.28892 | 0.41870 | 0.28407 | **8.26h** | | 0.29758 | 0.42636 | 0.29222 | 13.66h | | 0.30041 | 0.42857 | 0.29554 | **18.05h** | |
| Conv-KNRM_ListMLE | 0.25825 | 0.38841 | 0.25415 | 13.05h | | 0.25820 | 0.38982 | 0.25371 | 28.12h | | 0.24299 | 0.37543 | 0.23937 | 44.94h | | 0.24464 | 0.37705 | 0.24066 | 88.79h | |
| Conv-KNRM_nDCG | 0.20930 | 0.34968 | 0.20632 | 7.13h | | 0.26981 | 0.40201 | 0.26513 | 10.70h | | 0.28090 | 0.41167 | 0.27676 | **13.17h** | | 0.29054 | 0.41977 | 0.28578 | 19.43h | |
| Conv-KNRM_PoolRank | **0.30363¶** | **0.43270¶** | **0.29853¶** | 8.24h | 7 | **0.30814¶** | **0.43724¶** | **0.30344¶** | 13.66h | 5 | **0.31128¶** | **0.43954¶** | **0.30606¶** | 15.59h | 10 | **0.30588¶** | **0.43498¶** | **0.30104¶** | 22.77h | 20 |

PoolRank outperforms all pairwise learning approaches consistently and by a large margin in terms of MRR, nDCG and MAP. Also, PoolRank requires significantly less training time than the pairwise alternatives. Compared to other listwise learning schemes, PoolRank still yields the highest performance, especially for complex models such as ConvKNRM. ApproxNDCG performs reasonably well for KNRM model, while failing to yield consistent predictions for ConvKNRM model with different training document levels. The non-convex nature of the optimization task prevents metric-based methods such as ApproxNDCG from reaching global optima reliably. ListMLE shows its ability to enhance DRMM model, even though its DRMM performance is very close to PoolRank and ListNet. However, as model complexity increases and more documents are available for training, ListMLE pales in comparison with other algorithms with regard to convergence rate. ListNet converges very fast and produces competitive predictions with more documents available for training, *i.e.* the Top150/200, and it is less effective when fewer training documents are provided, *i.e.* the Top50.

PoolRank shows the best performance consistently across list sizes and models, while performance on Top50/100 candidates is on par with results on Top200 candidates. This indicates that PoolRank does not need too many training documents to achieve reliable results, and can therefore be memory efficient. Moreover, as model complexity increases (*i.e.* from DRMM to KNRM to ConvKNRM), PoolRank more effectively exploits the potential of neural retrieval models. The pooling techniques used in PoolRank only select the best document candidates for training, while extracting relative ordering signals based on min/max filtering much like a document-level dropout technique to accelerate model training. As a consequence, PoolRank consumes much less time while maintaining robustness and scalability.

### 4.4 Limited Training Data

In realistic settings, due to the high cost of data collection and human annotation, not all truly relevant documents will always be identified and available during model training. To further reflect on this perspective, we evaluate our PoolRank scheme on a proprietary news collection, which contains around 730k news articles covering various topics, with the restricted access to the relevant documents. We then construct 35 request queries based on this collection. Different from the generally shallowly judged MS MARCO collection, here, on average, 50 expert-judged relevant documents are pre-identified for each request. During model training, only at most two relevant documents are available for each query and the rest of pre-identified positive documents are marked as negative. But all the relevant documents are accessible for each request during model evaluation. Hence, this experiment further examines the effectiveness of the various loss functions when only limited positive candidates are available for model training and certain relevant documents are mislabeled. We train ConvKNRM, KNRM and DRMM for a fixed 30 epochs. To aid model training, BM25 is used to collect the top-5000 documents for each query. Pairwise/listwise training data is constructed in the same manner as before for MS MARCO. Table 2 reports the resulting model performance.

Table 2: Test Performance on proprietary news data.

| Model | DRMM | | | KNRM | | | ConvKNRM | | |
|---|---|---|---|---|---|---|---|---|---|
| | MRR | nDCG | MAP | MRR | nDCG | MAP | MRR | nDCG | MAP |
| Margin | 0.2452 | 0.3612 | 0.0623 | 0.9006 | 0.4614 | 0.1114 | 0.9059 | 0.5136 | 0.1613 |
| RankNet | 0.2836 | 0.3744 | 0.0703 | 0.8677 | 0.4542 | 0.1066 | 0.9145 | 0.5306 | 0.1883 |
| ListNet | 0.2807 | 0.3630 | 0.0665 | 0.8792 | 0.4651 | 0.1160 | 0.9224 | 0.5340 | 0.1761 |
| ListMLE | 0.1371 | 0.3415 | 0.0545 | 0.7076 | 0.4308 | 0.0978 | **0.9242** | 0.5003 | 0.1582 |
| ApproxNDCG | 0.3163 | **0.3802** | 0.0709 | 0.8963 | **0.4653** | 0.1153 | 0.9205 | 0.4991 | 0.1404 |
| PoolRank | **0.3337** | 0.3712 | **0.0732** | **0.9147** | 0.4624 | **0.1165** | 0.9147 | **0.5401** | **0.1928** |

Compared to all other pairwise/listwise learning methods, PoolRank yields consistently better performance across different metrics and neural retrieval models. Although ListMLE and ApproxNDCG offer competitive performance on certain metrics and specific models, they fail to yield even near average performance on other cases. This experiment validates the partial relevance assumption within PoolRank, which makes it more applicable in realistic information retrieval systems. For simplicity, we choose the same loss component coefficients and pooling window sizes as in the previous MS

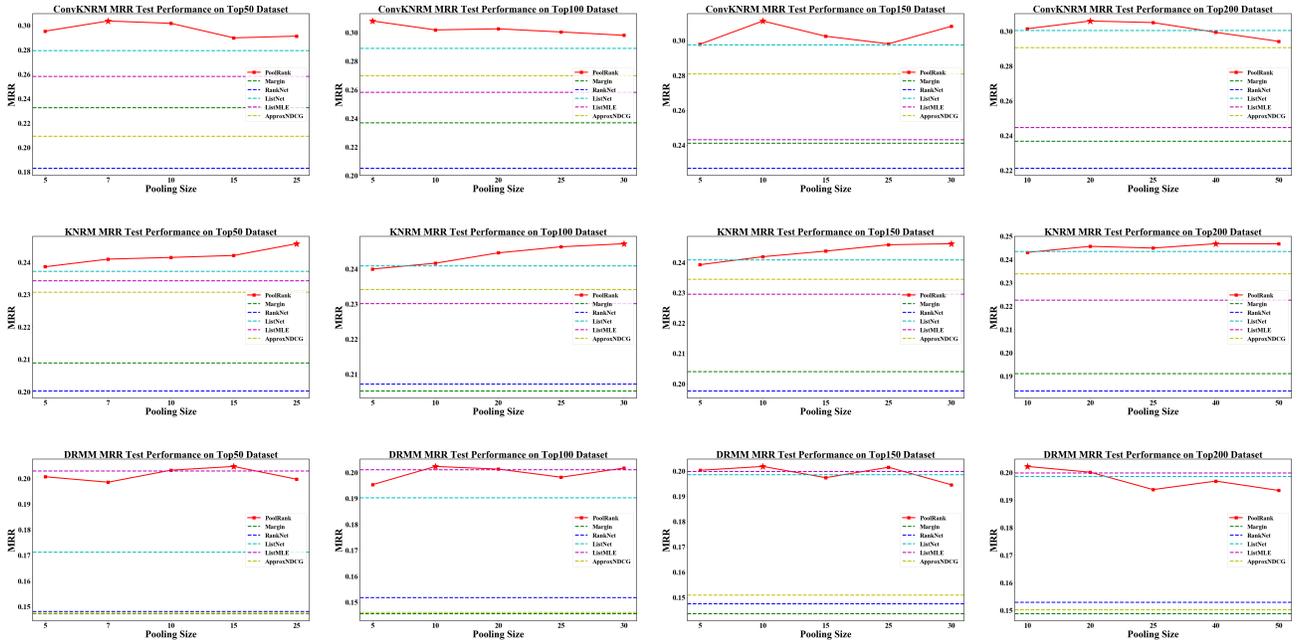

Figure 2: PoolRank MRR test performance on MS MARCO with different pooling sizes. ⋆ indicates the best performance. The five horizontal lines refer to pairwise margin ranking loss, RankNet, ListNet, ListMLE and ApproxNDCG, respectively.

MARCO experiment but believe that even more pronounced gains can be achieved by fine tuning those parameters based on each individual dataset.

## 5 AN ABLATION STUDY OF LOSS COMPONENTS

In all the experiments we showed above, PoolRank uses empirically tuned mixture weights $c_1 = 0.5, c_2 = 1, c_3 = 0.5, c_4 = 1$ based on validation dataset. PoolRank assumes that $L_{min}$ and $L_{min/max}$ create a ranking balance to stabilize the minimum scores to prevent them from becoming too small and $L_{min/max}$, $L_{max}$ and $L_{target}$ together establish a subtle ranking balance for the maximum scores to keep them being moderate. To further examine PoolRank's sensitivity to these two ranking balances, we apply PoolRank to ConvKNRM with window size 7 and list size 50 on the MS MARCO dataset with different selections of PoolRank loss components. Table 3 describes model test performance.[3] when one or more loss components are withheld (marked as "0").

We can observe that keeping the individual components of PoolRank balanced is essential to maintaining training effectiveness. If only a single component is used, the model invariably experiences dramatic performance declines. But if we choose two components, only a few combinations yield viable performance: 1) $L_{max}+L_{target}$ and 2) $L_{min} + L_{min/max}$, which validates our ranking balance assumption about minimum scores: $L_{min}$&$L_{min/max}$ maintain the minimum candidates at a reasonably lower score range. Furthermore, the loss with components of $L_{min/max}$, $L_{max}$ & $L_{target}$ shows the most significantly enhanced performance, compared to other three-component losses, which justifies these components' effectiveness at modeling the partial relevance of the maximum candidates.

[3]To preserve space, only the result of list size 50 is reported. The remaining list sizes have been investigated and follow the same trend indicated here.

Table 3: PoolRank loss component ablation on the MS MARCO with list size 50.

| $c_1$ | $c_2$ | $c_3$ | $c_4$ | MRR | nDCG | MAP | $c_1$ | $c_2$ | $c_3$ | $c_4$ | MRR | nDCG | MAP |
|---|---|---|---|---|---|---|---|---|---|---|---|---|---|
| 0 | 0 | 0 | 1 | 0.01568 | 0.13413 | 0.01538 | 1 | 0 | 1 | 0 | 0.18067 | 0.31866 | 0.17808 |
| 0 | 0 | 1 | 0 | 0.01299 | 0.13077 | 0.01276 | 1 | 1 | 0 | 0 | 0.28681 | 0.41881 | 0.28227 |
| 0 | 1 | 0 | 0 | 0.01468 | 0.13274 | 0.01447 | 0 | 1 | 1 | 1 | 0.29119 | 0.42148 | 0.28653 |
| 1 | 0 | 0 | 0 | 0.09831 | 0.24080 | 0.09620 | 1 | 0 | 1 | 1 | 0.27068 | 0.40421 | 0.26619 |
| 0 | 0 | 1 | 1 | 0.28211 | 0.41391 | 0.27810 | 1 | 1 | 0 | 1 | 0.14848 | 0.29062 | 0.14550 |
| 0 | 1 | 0 | 1 | 0.01563 | 0.13413 | 0.01531 | 1 | 1 | 1 | 0 | 0.17636 | 0.31656 | 0.17292 |
| 0 | 1 | 1 | 0 | 0.01226 | 0.13013 | 0.01202 | 1 | 1 | 1 | 1 | **0.29216** | **0.42406** | **0.28791** |
| 1 | 0 | 0 | 1 | 0.02149 | 0.14299 | 0.02097 | | | | | | | |

On the other hand, it is interesting to see that removing $L_{max}$ or $L_{target}$ can be detrimental to model performance, since the lack of these key components can break the ranking balance of maximum scores and compromise the final ranking results fatally. PoolRank consists of two coexisting ranking balance, $L_{min}, L_{min/max}$ for the minimum documents and $L_{min/max}, L_{max}$ & $L_{target}$ for the maximum candidates. Each ranking balance yields promising results. The subtle control over these two balance offer even more reliable and competitive performance as indicated in Table 1 and Table 3.

## 6 POOLING SIZE EFFECTS

In our listwise learning approach, we choose different pooling sizes to train neural retrieval models with different numbers of corresponding documents. Pooling size is a critical hyper-parameter in PoolRank. It controls the granularity of the information to extract within each candidate list. Compared to the traditional model structure based drop-out technique, pooling size in PoolRank can be regarded as the document-level drop-out rate. Within the pooling window of $p$ documents, only one candidate is selected for max-pooling layer, and another one for min-pooling layer. In the extreme case of $p = 1$, maximum candidates and minimum candidates are always the same. There is no distinct focus on the partial relevance problem. Thus, PoolRank can lead to the degenerative performance. For $p > 1$, the document drop-out rate is $(p-2)/p$. Moreover, pooling size also determines the distinguishability between maximum candidates and minimum candidates, since it controls how many documents to compare to select the corresponding maximum and minimum. Given the fixed candidate list size for listwise learning, too large pooling size yields more distinctive maximum candidates and minimum candidates, but it also selects much less documents for model training. On the other hand, too small pooling size may not be able to select proper maximum candidates that contain more partially relevant documents.

It's worth noting that different neural retrieval models show the different ability to distinguish different documents based on the different extracted semantic features. Better models usually can detect linguistic nuance even from vast documents and yield better ranking performance. Therefore, we show how performance evolves for various models with respect to pooling size. Since there are usually only one relevant document for the given query on MS MARCO dataset, ranking metrics, MRR, nDCG and MAP, are approximately proportional to each other and show similar trends. Hence, in Figure 6, we only plot the MRR test performance for each neural retrieval model with different pooling sizes. The dashed lines show the performance of other baseline loss functions we tested, pairwise margin loss, RankNet, ListNet, ListMLE and ApproxNDCG.

Comparing the results of pairwise learning, our PoolRank approach performs better consistently with different pooling sizes in general. When there are more document candidates available, the general performance gap between PoolRank and pairwise learning increases. Although ListNet tends to require larger number of candidates to reach comparable performance, PoolRank with different pooling sizes produces consistent, competitive and reliable performance across different sub-datasets. This means less candidate list is sufficient for PoolRank to reach model potentials. In the realistic production setting, smaller list size means less computing memory requirement for listwise learning, which can encourage more researchers to employ listwise learning to exploit neural model potential.

Specifically, complex models such as ConvKNRM prefer the relatively small pooling size. Due to its strong ability to distinguish documents, small pooling sizes lead to more candidates for model training, while the contrast between maximum/minimum candidates can still be detected. On the other hand, simpler models such as KNRM prefer larger pooling size. As pooling size increases, the pooling-based document difference becomes more useful and reliable for KNRM. Also, simpler models generally require less data for the effective model training. Hence, less selected pooling-based candidates can still be sufficient for KNRM learning. DRMM relies on the fixed word embedding and pre-calculated semantic signals, such as matching histogram mapping. The number of parameters of DRMM is significantly less than even KNRM model. Model learning is all about feeding loss-based update to improve the configuration of the internal neural network parameters. The simplicity of DRMM limits its potential to generate comparable performance. Thus, PoolRank with various pooling sizes leads to the fluctuating results.

## 7 CONCLUSION

In this paper, we present PoolRank, a pooling-based listwise learning framework that is compatible with arbitrary neural retrieval models. Relevance annotation can be a costly, labor-heavy and time-consuming task for human judges. The false negative rate in many shallowly judged IR collections can potentially be high and may therefore mislead derived neural models. To address this problem, PoolRank introduces the concept of ranking balance and the direct consideration of partial relevance via the classic pooling technique. Within PoolRank, a min-pooling layer selects the most obviously non-relevant candidates at the local level and a max-pooling layer chooses the potential partially relevant, yet non-relevant-labeled candidates carefully based on the prediction comparison among the local candidates. Inspired by mechanical equilibrium in physics, the four components of PoolRank then preserve an effective ranking balance for the selected minimum/maximum candidates. For the minimum candidates, $L_{min}$ & $L_{min/max}$ prevent their ranking scores from being too small while still remaining at the lower range. For the maximum candidates, $L_{min/max}$, $L_{max}$ & $L_{target}$ together build a critical balance to keep their scores being moderate and reflect their partial relevance directly on the final ranking prediction. In our experiments on MS MARCO, we show PoolRank to outperform all traditional pairwise/listwise methods for three different neural retrieval models and four scenarios with strongly varying numbers of available candidate documents per query. With the special design in the proprietary news collection experiment, only certain limited relevant documents are identified for model training. PoolRank still produces the consistent, competitive and reliable performance across different neural retrieval models.

There are several promising directions for future inquiry building on this work. Most notably, PoolRank is the first listwise ranking loss to make use of pooling techniques. To further improve on its performance, we are curious to integrate these pooling-based approaches to existing listwise learning methods, harnessing their joint strengths.

## ACKNOWLEDGMENTS
Acknowledgements blinded for anonymous review.